\documentclass[sigconf,screen,nonacm]{acmart}

\AtBeginDocument{%
  \providecommand\BibTeX{{%
    \normalfont B\kern-0.5em{\scshape i\kern-0.25em b}\kern-0.8em\TeX}}}

\setcopyright{acmcopyright}
\copyrightyear{2023}
\acmYear{2023}
\acmDOI{XXXXXXX.XXXXXXX}

\acmConference[ICAIF '23]{4th ACM International Conference on AI in Finance}{November 27--29, 2023}{New York City, NY}

\usepackage{enumitem}%
\usepackage{amsmath}
\usepackage{booktabs}
\usepackage{adjustbox}
\usepackage{subcaption}
\newlist{todolist}{itemize}{2}
\setlist[todolist]{label=$\square$}
\usepackage{pifont}

\usepackage{svg}

\begin{document}

\title{Generative AI for End-to-End Limit Order Book Modelling}
\subtitle{A Token-Level Autoregressive Generative Model of Message Flow Using a Deep State Space Network}

\author{Peer Nagy}
\email{peer.nagy@eng.ox.ac.uk}
\affiliation{%
  \institution{Oxford-Man Institute of Quantitative Finance, University of Oxford}
  \country{UK}
}

\author{Sascha Frey}
\affiliation{
    \institution{Department of Computer Science, University of Oxford}
    \country{UK}
}

\author{Silvia Sapora}
\affiliation{
    \institution{Foerster Lab for AI Research, University of Oxford}
    \country{UK}
}

\author{Kang Li}
\affiliation{
    \institution{Department of Statistics, University of Oxford}
    \country{UK}
}

\author{Anisoara Calinescu}
\affiliation{
    \institution{Department of Computer Science, University of Oxford \& Alan Turing Institute}
    \country{UK}
}

\author{Stefan Zohren}
\affiliation{
    \institution{Oxford-Man Institute of Quantitative Finance, University of Oxford \& \\ Man Group}
    \country{UK}
}

\author{Jakob Foerster}
\affiliation{
    \institution{Foerster Lab for AI Research, University of Oxford}
    \country{UK}
}

\renewcommand{\shortauthors}{Nagy et al.}

\begin{abstract}
  Developing a generative model of realistic order flow in financial markets is a challenging open problem, with numerous applications for market participants. Addressing this, we propose the first end-to-end autoregressive generative model that generates tokenized limit order book (LOB) messages. These messages are interpreted by a \emph{Jax-LOB} simulator, which updates the LOB state. To handle long sequences efficiently, the model employs \emph{simplified structured state-space layers} to process sequences of order book states and tokenized messages. Using LOBSTER data of NASDAQ equity LOBs, we develop a custom tokenizer for message data, converting groups of successive digits to tokens, similar to tokenization in large language models. Out-of-sample results show promising performance in approximating the data distribution, as evidenced by low model \emph{perplexity}. Furthermore, the mid-price returns calculated from the generated order flow exhibit a significant correlation with the data, indicating impressive conditional forecast performance. Due to the granularity of generated data, and the accuracy of the model, it offers new application areas for future work beyond forecasting, e.g. acting as a world model in high-frequency financial reinforcement learning applications. Overall, our results invite the use and extension of the model in the direction of autoregressive large financial models for the generation of high-frequency financial data and we commit to open-sourcing our code to facilitate future research.
\end{abstract}



\settopmatter{printfolios=true}
\maketitle

\section{Introduction}

Ever since OpenAI opened access to ChatGPT, generative large language models (LLMs) have skyrocketed in popularity. Part of their appeal is due to their impressive few-shot learning and in-context learning abilities \cite{brown2020language}. Besides LLMs, generative diffusion models have had a similar rise in popularity for image generation \cite{croitoru2023diffusion}. Financial applications similarly benefit from generative models for various applications, from data augmentation \cite{naritomi2020data, liu2022synthetic}, over anomaly detection \cite{henry2019generative, geiger2020tadgan}, to forecasting \cite{zhou2018stock}.
Most current financial machine learning approaches employ training paradigms based on generative adversarial networks (GANs) \cite{goodfellow2014generative, eckerli2021generative}, which usually generate series of price returns directly. Just a few years ago, \citet{takahashi2019modeling} stated that ``Building auto-regressive models of financial time-series meets insurmountable difficulties.'' Only very recently has there been work on bottom-up generators of limit order book (LOB) market micro-structure data \cite{hultin2023generative}. Our paper proposes to take the next step towards more powerful autoregressive financial micro-structure generative models, which do not suffer from problems commonly encountered with GANs, such as mode collapse \cite{bau2019seeing, zhang2018convergence}. Autoregressive models enable generating sequences of variable length and allow for improved interpretability, as the model learns conditional distributions of the next sequence token by design. The trained model, in conjunction with a replay simulator \cite{frey2023jaxlob}, which processes generated messages, is used to generate a realistic order flow. Such a conditional simulation model constitutes an interactive world model, which, in principle, can be used in downstream tasks, such as trading, order execution, or market making, potentially using a reinforcement learning algorithm \cite{coletta2023conditional}. We leave these applications to future work.

We propose the use of a \emph{deep state space model} composed of \emph{simplified structured state space} layers for sequence modeling (S5) \cite{smith2022simplified}, which are computationally efficient and excel at learning long-range dependencies. Recognizing the similarity between sequences of order book messages and natural language, our model learns the conditional data distribution of \emph{tokenized} message sequences in a cross-entropy minimization task, accurately predicting the next token in the sequence. A token in a natural language task corresponds to a part of a word or a sequence of successive digits, whereas here it is the latter. The autoregressive setup has numerous virtues, compared to GANs, such as model scalability -- as evidenced by LLMs -- but also wins in model interpretability, as the model defines a distribution over all message sequences of arbitrary length. Challenges include modeling extremely long sequences (>10,000 tokens), as we need to represent \emph{sequences of sequences}, since each message is itself a sequence of tokens; and long-range data referentiality.

To our knowledge, this is the first paper to propose an autoregressive end-to-end generative model for micro-structure messages. Furthermore, we develop a tokenizer, converting LOB messages to a finite vocabulary of tokens. Model performance is quantified by calculating \emph{perplexity} scores and by evaluating conditional distributions of generated data in an inference loop (see section \ref{sec:results}) with a custom error correction mechanism (see section \ref{sec:model_inf}). We utilize the Jax framework \cite{jax2018github} for hardware acceleration of both the model and simulator, accelerating training and model inference.

\section{Related Literature} \label{sec:related}

The generation of synthetic financial data and LOBs is an active topic of research, with a variety of approaches employed, including deep learning methods \cite{assefa2020generating, SEZER2020106181}.
Traditional methods focus on computational statistical methods to generate probabilities of LOB events but are often limited, as they make strong assumptions \cite{cont2010stochastic} and are not accurate enough to be used in many practical applications.
Another approach is agent-based modeling (ABM). In ABMs, simulations are used to understand how the interaction of autonomous agents leads to aggregate statistics and emergent behavior in a system \cite{korczak2017deep, byrd2020abides}.
Recently, generative neural networks have been used as black-box models of the dynamics of LOBs \cite{coletta2021towards, coletta2022learning, hultin2023generative}. Recent work has focused on the use of LSTMs \cite{rumelhart1987learning, hultin2023generative} and generative adversarial networks (GANs) \cite{goodfellow2014generative}, such as \citet{coletta2021towards, coletta2022learning}. Many variations of GANs have been developed, including autoregressive implicit quantile networks \cite{ostrovski2018autoregressive}, ExGAN \cite{bhatia2021exgan}, TimeGAN \cite{NEURIPS2019_c9efe5f2} and CGAN \cite{DBLP:journals/corr/MirzaO14}, making GANs more easily applicable to financial data.

Examples of GAN's applications to probabilistic forecasting of financial time series include Tail-GAN \cite{cont2022tail} and FinGAN \cite{vuletic2023fin}. Both use custom economics-driven loss functions, making them better suited for financial applications. \citet{vuletic2023fin} adapt FOR-GAN \cite{Koochali_2019}, a combination of CGANs and RNNs used in probabilistic forecasting. 
\citet{cont2022tail} simulate multivariate prices with the goal of producing accurate tail risk statistics for a set of benchmark strategies. To do this, they combine GANs with Principal Component Analysis (PCA), allowing the architecture to scale effectively to a large number of assets. Their loss function is a bi-level optimization equivalent to a max-min game, where the discriminator's goal is to predict Value-at-Risk and Expected Shortfall. %

An attempt at generating LOB messages directly on a granular level is \citet{doi:10.1080/14697688.2023.2205583}. They model the LOB as a continuous-time Markov chain with volumes across price levels defining the state, extending the model of \citet{cont2010stochastic}. For each state transition, every feature (event type, order size, and price) is modeled separately, using a dedicated RNN per feature. The joint probability is then obtained from the product of the individual conditional probabilities. Instead of tokenizing messages, large numerical values -- such as order sizes or timestamps -- are binned, leading to a loss of precision. In contrast, our approach models the full level-3 representation, referencing individual orders, rather than the level-2 aggregation.

\section{Background} \label{sec:background}

\subsection{Limit Order Books (LOBs)} \label{sec:background_lobs}
The LOB contains a set of all unmatched limit orders submitted to an exchange and a mechanism by which incoming orders are matched \cite{gould_limit_2013}. In a price-time priority book, limit orders are ordered, first by their price, and second by their arrival time. For buy orders (bids), the orders with the highest price are prioritized, and for sell orders (asks) the lowest prices are. When an incoming order ``crosses the spread'', i.e. accepts a lower (higher) price than the best bid (ask), the incoming order is matched with the existing orders according to price-time priority. Upon matching two orders, a trade occurs, and the ownership of the underlying security is transferred. The evolution of the state of the LOB can be wholly reconstructed given an initial book state and the total set of arriving order messages in a time interval. LOB messages types are described in Section \ref{sec:data}.

\subsection{State-Space Models} \label{sec:background_s5}
Recently, transformers \cite{vaswani2017attention} have been the most successful and widely used approach to long-range sequence modeling \cite{Rombach_2022_CVPR, jumper2021highly, chowdhery2022palm}, despite their quadratic complexity $O(L^2)$ in sequence length $L$. The S4 architecture \cite{gu2021efficiently} (and subsequent variants S4D \cite{gu2022parameterization} and S5 \cite{smith2022simplified}) achieve state-of-the-art performance in the \emph{Long Range Arena} task \cite{tay2020long}, which is designed to measure long-range reasoning abilities, while maintaining linear complexity in $L$ at inference time. The S4 architecture employs the state-space model (SSM), commonly used in control theory. The SSM is defined as:

\begin{equation}
\begin{aligned}
\mathbf{x'}(t) &= A\mathbf{x}(t) + B\mathbf{u}(t), \\
\mathbf{y}(t) &= C\mathbf{x}(t) + D\mathbf{u}(t). 
\end{aligned}
\end{equation}

The state vector $x(t)$ denotes the current state of the system and $u(t)$ the input vector, which is the set of variables affecting $x(t)$.
Combined with deep learning and the HiPPO framework \cite{gu2020hippo}, matrices $A$, $B$, $C$ and $D$ can be learned through standard gradient descent to achieve high performance. While the SSM equations are defined in continuous-time, they can be easily discretized using either the bilinear method \cite{tustin1947method}, or zero-order hold (ZOH), using a fixed time step.
S4's features allow it to be extremely computationally efficient: it is easily parallelizable during training by ``unrolling'' it as a convolution using fast Fourier transform (FFT), and has $O(H^2)$ complexity per step at inference time (where $H$ is the hidden dimension).
The S5 model further improves on its predecessor by employing a parallel scan operation to compute the hidden state each step and using matrix multiplication instead of applying a convolutional kernel. This is facilitated by S5's use of one multi-input, multi-output SSM instead of the many single-input, single-output SSMs in S4, and the approximate diagonalization of the HiPPO matrix. This allows the model to  maintain linear complexity over sequence length while also being able to model time sequences with varying sampling time steps \cite{smith2022simplified}.

\subsection{Autoregressive Models}

An autoregressive generative model, for example an LLM, defines a (conditional) probability distribution over entire sequences of tokens. Model quality can thereby be measured by evaluating the probability the model assigns to the test data. One such statistic that is often applied in NLP is perplexity (PPL) \cite{chen1998evaluation}. PPL  measures the expected per-token surprisal by the data. It is the exponential of the per-token cross-entropy loss:

\begin{equation}
\begin{aligned}
    PPL &= \exp \left\{ \frac{1}{N} \sum_{\{\mathbf{m}, \mathbf{b}, y\}} - \log p(y | \mathbf{m}, \mathbf{b}) \right\} \\
        &= \exp \left\{ \frac{1}{N} \sum_{\{\mathbf{m}, \mathbf{b}, y\}} - \log \hat{y}_{\{i=y\}}  \right\},
\end{aligned}
\end{equation}
where $\hat{y_i} = p(y_i | \mathbf{m}, \mathbf{b})$ denotes the conditional probability of token $i$ under the model. Perplexity for individual sequences is calculated as the exponential of the mean loss over autoregressive one-step-ahead predictions of all tokens of that sequence. This is because log probabilities of sequences are conditionally separable:

\begin{equation*}
\begin{aligned}
\log\ p(y_1 y_2 \ldots y_n) &= log\ p(y_1) + log\ p(y_2|y_1) + log\ p(y_3|y2,y1) \\
                            &+ \ldots + log\ p(y_n|y_{n-1},\ldots, y_1).
\end{aligned}
\end{equation*}

\section{Data} \label{sec:data}

We use the LOBSTER data of LOBs of NASDAQ cash equities \cite{huang2011lobster}. In particular, we train and evaluate our model separately on data from Alphabet (GOOG) and Intel (INTC). For each stock, we use 102 days of training data (1 July 2022 to 11 November 2022), and 12 days of validation (28 November to 13 December 2022) and test data (14 December to 30 December 2022). These two stocks represent examples of small-tick (GOOG) and large-tick LOBs, which exhibit different dynamics \cite{eisler2012price}. Large-tick stocks, those where the tick size of \$0.01 is large relative to the stock price, tend to have less sparse LOBs and a more constant spread over time. We only use data during regular NASDAQ trading times, on working days between 09:30 and 16:00 US East Coast time. We utilize level-3 messages pertaining to the best $l$ price levels recorded at any given moment in time. In contrast to level-2 data, where only aggregate open volumes are recorded per price level, level-3 messages allow for a full-fidelity reconstruction of LOB dynamics within certain constraints. For each new message, the data set also contains a snapshot of prices and volumes at the best $l$ price levels, for both buy orders (bids) and sell orders (asks). We use $l=10$ levels of LOB data, which is a common threshold for full LOB investigations \cite{cont2021cross}.

LOBSTER data contains 7 message types: new limit orders, partial order cancellations, full order deletions, visible order executions, hidden order executions, auction trades, and trading halts. However, we only use the first 4 types, ignoring hidden orders, which do not affect the visible dynamics of the LOB, and trading halts. Auction trades only occur outside regular daily trading times, and are therefore not contained in our data. Each message has 6 constituting elements: a timestamp in nanoseconds after midnight, the event type (1-4), the order ID, the order size, the price, and the trading direction (buy or sell). An example order is shown in Figure \ref{fig:encoding}.

\subsection{Pre-processing}

The data is pre-processed to make it better suited for a deep learning task, such as generative modeling. However, our choice of data representation is general enough for other machine learning tasks. Our proposed model architecture uses both order flow information, in the form of messages, and sequences of level-2 order book states. While some models use LOB data in a \emph{price-volume} representation, e.g. DeepLOB \cite{zhang2019deeplob}, transforming the data into a more stationary representation can considerably improve performance \cite{kolm2021deep}. We propose and use a sparse representation of liquidity in the book as $P$ separate volume features around the mid-price, coupled with one feature representing mid-price changes from the previous observation. This results in a $P+1$ dimensional vector at each point in time, which is sparse when the LOB's price levels are sparsely populated. This representation fixes a dollar-range \emph{volume image}, which also preserves price levels in the price range without any volume. Empty levels can carry important information, such as the size of the bid-ask spread, or the shape of book volume at deeper levels, which are usually squashed in other data representations. Furthermore, our representation is especially useful for a generative model, since the model can place orders at a new price level without changing the absolute price a specific book feature refers to. Since we do not generate order book states directly, but message sequences, there is no need to tokenize the book data. Instead, these can be input directly into the model as continuous features. %

Before messages are encoded as token sequences, we convert them to a more stationary data representation and a finite token vocabulary. 
After filtering the data to the relevant order types, prices are converted from dollar values to ticks (cents) from the previous mid-price. In case the mid-price lies between valid ticks, it is rounded down to the next tick. Modified prices can thus be represented as integers and become more stationary over time. We then truncate modified prices below -999 and above 999 ticks around the mid-price to enable encoding the data with a finite and fixed vocabulary. Similarly, extremely rare individual large orders are truncated to an order size of 9999. In practice, these thresholds affect less than 0.1\% of the data.

Besides new limit orders, all other message types are referential, as they subtract liquidity from an existing open order in the book. This might be obvious for partial order cancellations or full deletions, but it is modeled analogously for order executions. Execution messages must therefore refer to the order ID of the highest priority bid or ask order in the book. A challenge arises in encoding this referentiality in the message data, since numeric order IDs are a poor choice, due to their arbitrary nature and non-stationarity. As an alternative, we append information about the referenced original message as additional features to all messages. These fields are the original modified price, size, and timestamp, which we use to identify the referenced limit order.\footnote{In the rare case of exactly identical orders, recorded at the same nanosecond, we choose the last order in the list.} Since arrival times are monotonically increasing, and thus a difficult distribution to generate from, we also add inter-arrival times between messages as additional features. Finally, we reorder the message features so that longer fields, corresponding to higher-entropy token distributions, occur later in the message and can thus condition on previous fields in our target task of left-to-right ``causal'' prediction. The order of features is: event type, direction, price, size, inter-arrival time $\Delta t$, arrival time; followed by the reference fields: price, size, and time of the original message. In the case of a new limit order, the reference fields receive an NA value.

\subsection{Encoding} \label{sec:encoding}

\begin{figure}[htbp]
  \centering
  \includegraphics[width=\linewidth]{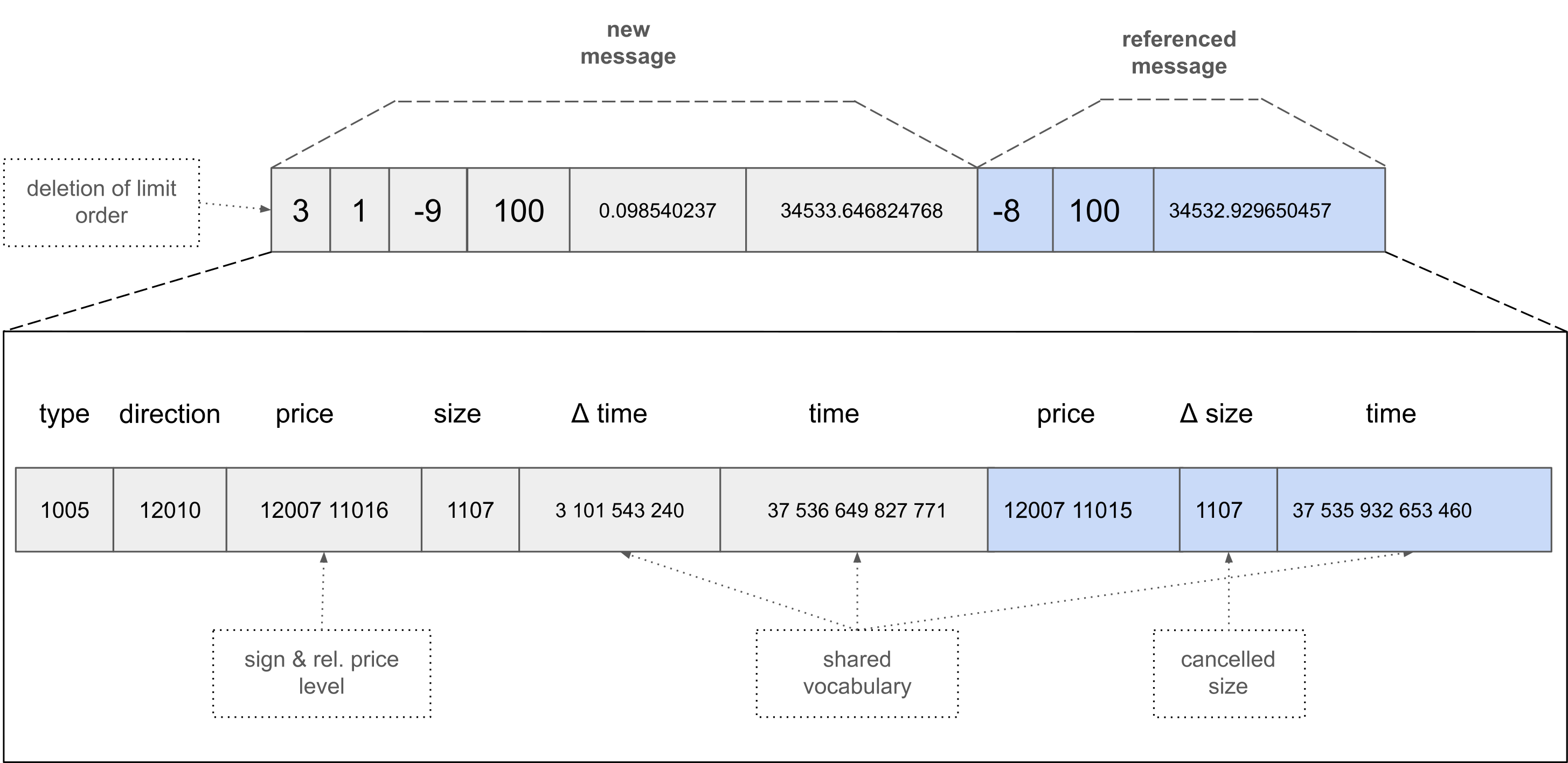}
  \caption{Schematic of message tokenization of a limit order deletion message. Each field of the pre-processed message at the top corresponds to a sequence of tokens in the encoded message at the bottom. The grey message part corresponds to the new message arriving at the exchange, while the blue part encodes the referenced message to be deleted.}
  \label{fig:encoding}
\end{figure}

Analogously to a generative language model, we propose a \textit{token-based encoding} scheme for LOB messages. This allows the model to be trained using cross-entropy loss on flattened token sequences so that it learns the conditional distribution over a target token. Figure \ref{fig:encoding} illustrates the encoding mechanism of a pre-processed order deletion message. Here, each field of the pre-processed message corresponds to a sequence of tokens in the encoded message. The message vocabulary contains 12011 distinct tokens, which are represented as integer values. Event type, direction, and size fields are each encoded with a single token, while prices are split into two, the first corresponding to the sign (above or below the mid-price) and the second to the tick distance from the mid-price. A difference to language encoders is that we use non-overlapping token ranges for some fields. For example, even though event type and price level might have the same raw numerical value, they are encoded using different tokens to make effective use of our a priori knowledge of their semantic difference. Due to their similarity, arrival times and inter-arrival times ($\Delta t$) share the same vocabulary and are tokenized in groups of 3 digits. A timestamp with nanosecond precision (15 digits) thus corresponds to a 5-token sequence and the 9 fields of the pre-processed message are converted to 22 tokens.
While we conduct all data pre-processing for our entire data set before model training and inference, message encoding and decoding are done dynamically when loading the data. Dynamically decoding messages is required for generated messages to be submitted to the simulator during model inference. This is done computationally efficiently by relying on just-in-time (JIT) compilation on hardware accelerators (such as GPUs) using the Jax framework \cite{jax2018github}.

\section{Model} \label{sec:model}

\subsection{Model Architecture}

The model architecture uses a deep network of \emph{simplified structured state-space layers} (S5) \cite{smith2022simplified} (see section \ref{sec:background_s5}). Model inputs are flattened sequences of $n$ tokenized LOB messages $\mathbf{m} \in \mathcal{V}^{22n}$ and of $n$ \emph{volume images} of the level-2 LOB states $\mathbf{b} \in \mathcal{B}^n$, where $\mathcal{V} \subset \mathbb{N}$ is the token vocabulary and $\mathcal{B}$ the space of transformed book snapshots containing P volumes $\in \mathbb{R}_{+}$ and the previous mid-price change $\in \mathbb{Z}$.
This way, message $m_i$ at sequence location $i \in [0,n-1]$ acts upon order book state $b_i$ and transitions it to $b_{i+1}$. So, for each book state, the model receives the corresponding successor message.

Similarly to masked language modeling \cite{kenton2019bert}, during training, we mask a random token in the last message, using a $MSK$ token, and replace all tokens to the right of this with $HID$ tokens, so the model does not condition on those fields. Time tokens of new messages are not predicted but instead calculated from generated inter-arrival times $\Delta t$. Therefore, these are not selected to be masked during training. Time tokens do however remain part of input sequences and time tokens of referenced messages \emph{are} prediction targets as these are essential in identifying original messages. These are easier targets, as they are already part of the message sequence when the referenced limit order had been submitted, which is often contained in the input sequence.

We define the model $f_{\mathbf{\boldsymbol{\theta}}} : (\mathbf{m}, \mathbf{b}) \mapsto \mathbf{\hat{y}}$, parametrised by $\boldsymbol{\theta}$, as mapping a message sequence $\mathbf{m}$ and a book sequence $\mathbf{b}$ to a vector of logits $\mathbf{\hat{y}} \in \mathbb{R}^v$ where $v=|\mathcal{V}|$ is the size of the token vocabulary. By requiring that $\sum_{i=1}^v \exp(\hat{y}_i) = 1$ the model defines a distribution over $\mathcal{V}$, which learns the distribution of the masked token, conditional on the input sequences.
To achieve this, the model parameters $\boldsymbol{\theta}$ are trained by minimizing cross-entropy loss  

\begin{equation}
\begin{aligned}
    &\min_{\boldsymbol{\theta}} \frac{1}{N} \sum_{\{\mathbf{m}, \mathbf{b}, y\}} L(y, \mathbf{\hat{y}}) 
    = \min_{\boldsymbol{\theta}} \frac{1}{N} \sum_{\{\mathbf{m}, \mathbf{b}, y\}} -\log \hat{y}_{\{i=y\}} , \\
    &\text{where} \quad \mathbf{\hat{y}} = f_{\boldsymbol{\theta}}(\mathbf{m}, \mathbf{b}),
\end{aligned}
\end{equation}
over $N$ tuples of training data $\{\mathbf{m}, \mathbf{b}, y\}$ using gradient descent with the Adam \cite{kingma2014adam} optimizer. The scalar $y \in \mathcal{V}$ denotes the target token from the data. In practice, tokens are one-hot encoded, so that the model learns an input embedding.

\begin{figure}[htbp]
  \centering
  \includegraphics[width=\linewidth]{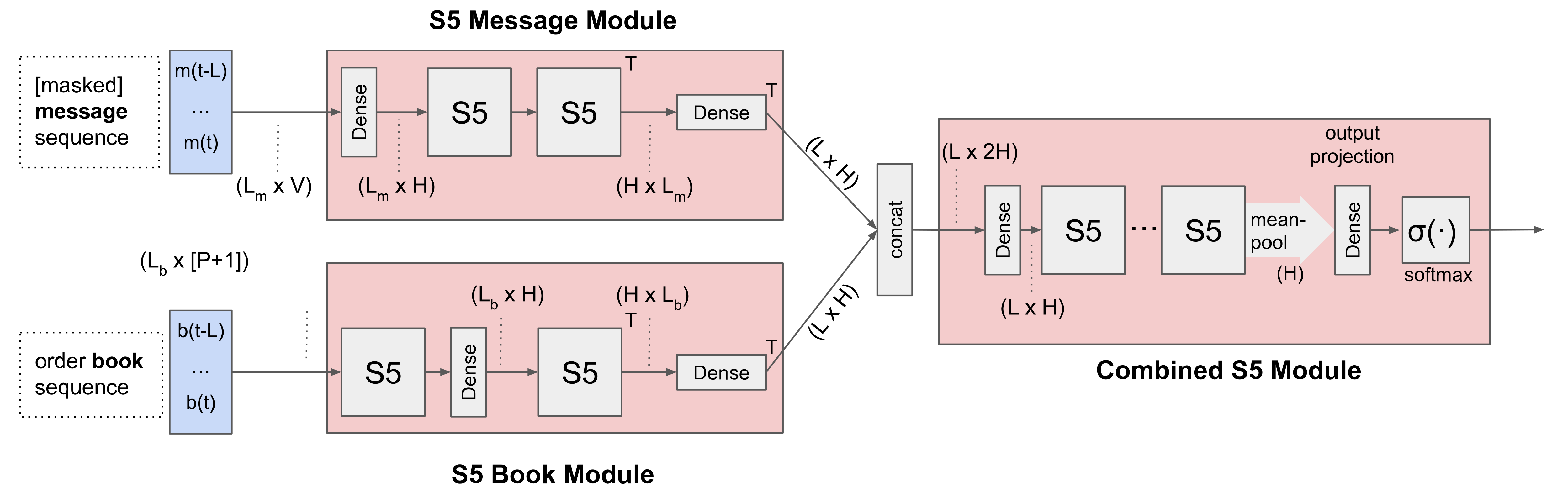}
  \caption{Schematic of model architecture. Book sequences and flattened tokenized message sequences are initially processed separately before being processed by a combined S5 module (using 6 S5 layers, resulting in $\sim6.3x10^{6}$ parameters). Sequence outputs from the last S5 layer are averaged along the sequence dimension to produce token logits.}
  \label{fig:model}
\end{figure}

The model architecture is shown in Figure \ref{fig:model}. The network has two separate input branches, one receiving the masked message sequences, and the other the book state sequences. Both are processed separately for a few layers, combining S5 with dense layers, before projecting both sequences to a common sequence length $L$. The concatenated sequences are then further processed by a stacked block of S5 layers, before projecting the output to $v$ output neurons. The message branch starts with a linear embedding layer, projecting each one-hot token vector to the model's hidden dimension $H$, followed by S5 layers. The book branch, on the other hand, first passes each $(P+1)$-dimensional observation through an S5 layer before also projecting it to the embedding dimension $H$.

We trained the model on input sequences of $n=500$ messages and LOB states, corresponding to encoded sequence lengths of 11,000 tokens for the messages (22 tokens per message), and 500 observations for the book states as these are not tokenized. To vary the exact input sequences and $MSK$ token location, in every training epoch, a random number of observations -- between 0 and $n-1$ -- are skipped from the start of each day.

\subsection{Inference: Combining Model and Simulator} \label{sec:model_inf}

We learned that training and validation loss drastically improves when using book data in addition to message sequences. To reconstruct the state of the LOB solely from order flow, extremely long message sequences need to be combined non-linearly, which is why both have been found to contain orthogonal information in prediction tasks \cite{zhang2021deep}. While it is easy to combine both data sources during training, we require a mechanism to generate new book states during multi-step autoregressive inference.

Our solution is a LOB simulator, which takes the most recent LOB state and applies the generated message according to LOB matching rules. To ensure proper processing by the simulator, the generated message is initially decoded and subsequently passed through an error correction mechanism. The simulator is therefore an essential component of the inference pipeline. We utilize the novel LOB simulator, introduced in \citet{frey2023jaxlob}, implemented in Jax \cite{jax2018github}, and end-to-end just-in-time (JIT) compiled for GPU.

\begin{figure}[htbp]
  \centering
  \includegraphics[width=\linewidth]{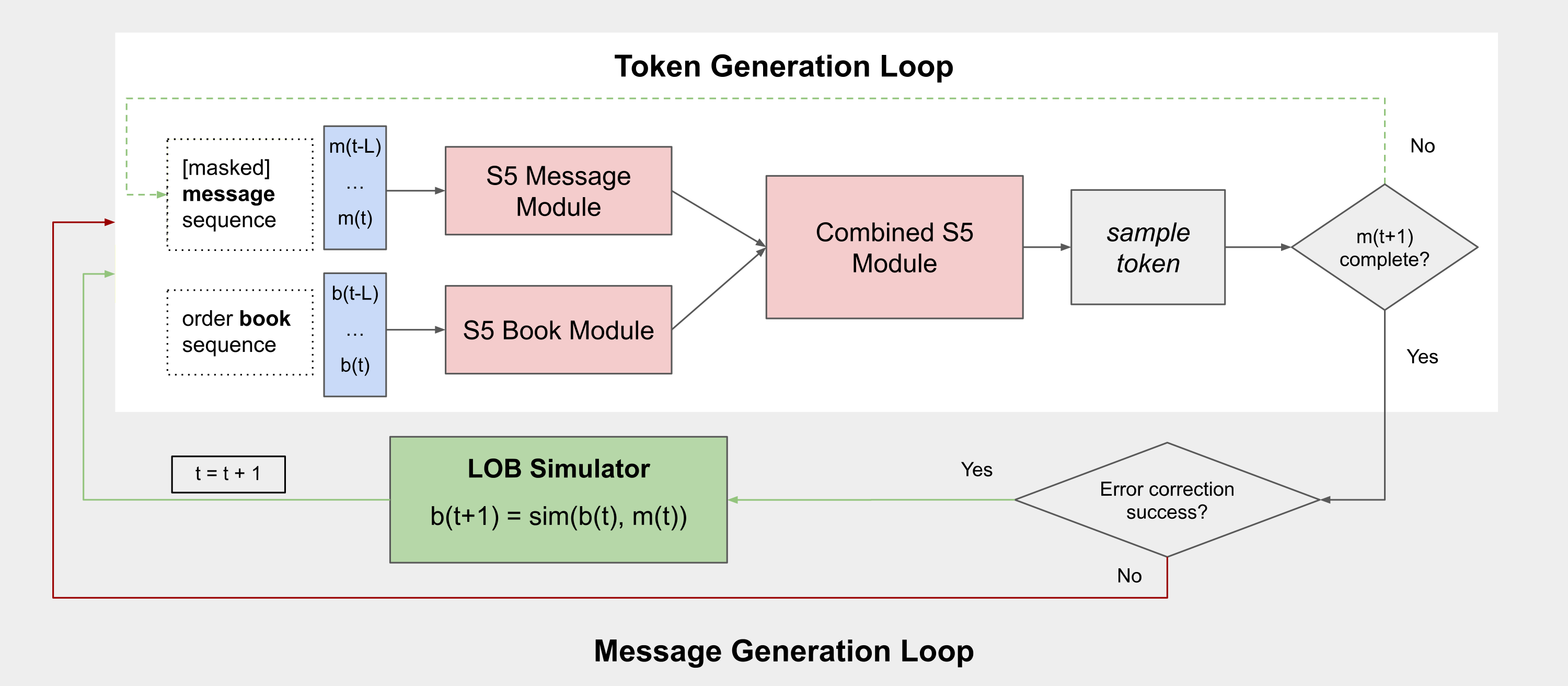}
  \caption{Inference Loop. Tokens are generated autoregressively from model logits until a message is complete (Token Generation Loop). Completed messages are error-corrected and input to the LOB simulator. The updated book state and the generated message are then added to the input sequences to start generation at $t+1$. (Message Generation Loop)}
  \label{fig:inference_loop}
\end{figure}

The inference pipeline is illustrated in Figure \ref{fig:inference_loop}. It starts with the model receiving a tokenized message sequence, where the first token of the last message is set to $MSK$, and the remainder to $HID$. Tokens are then sampled autoregressively left-to-right using the \emph{softmax} over output logits. Once all tokens constituting $\Delta t$ have been sampled, the message's arrival time $t$ is calculated by adding $\Delta t$ to the previous message's time stamp. The new arrival time is then tokenized and inserted into the message sequence. During sampling, we restrict the distribution to syntactically valid tokens for the currently masked field but otherwise sample proportionately to the predicted token scores without truncation or beam-search.

After initializing the simulator with a LOB state, the message sequence is then replayed to advance the state to the current time step. This is necessary as the book states are represented as a level-2 image, which means that we only have the aggregate volume of each individual order at a price level. This initial volume is represented by the simulator as a single order. By replaying the message sequence, we can thereby recreate a partial level-3 representation, which is required to replay referential orders.

The role of the error correction procedure is to correct the occasionally hallucinated message reference components, which do not exist in the data and the simulator. As new limit orders do not require a reference component, this is only done for cancellations, deletions, and order executions. Past limit orders in the sequence, which are still in the LOB, are first searched for an order direction, price, size, and time matching the generated reference. If this matching should fail, the search is repeated excluding the time field as this field exhibits the highest error rates. Should there still be no satisfactory match, which would happen if a message references order flow before the start of the input sequence, cancellations are applied to the initial volume, if the simulator has any left at the correct price level. As executions of multiple order blocks at the same time are modeled as separate events, correct order executions are easier to guarantee. There is only a single referenced candidate order on each side of the book, namely the limit order at the best price with the earliest arrival time. 

\section{Results} \label{sec:results}

The primary objective of a generative model is to produce data that closely approximates the target distribution. We propose to evaluate model performance in three distinct ways.
\begin{enumerate}[leftmargin=*]
\item We compare various unconditional marginal distributions produced by the model with the corresponding data distributions. \item Next, to evaluate the model's capacity to match conditional distributions, correlations between generated mid-price returns and realized returns are calculated. Results indicate that the model's significant forecasting horizon is competitive with deep learning models, trained explicitly to forecast mid-prices \cite{zhang2019deeplob}.
\item Finally, as a succinct measure of model performance across the entire distribution, we calculate \emph{perplexity} scores, which are common in the evaluation of large language models (LLMs) \cite{chen1998evaluation}.
\end{enumerate}

For evaluation purposes, we sample 1000 random test sequences, each comprising 600 messages and corresponding LOB states, which temporally follow the training and validation period. From each sequence, we extract the first $n=500$ observations between time steps $t-n+1$ and $t$ as the model input. The model generates the succeeding 100 messages from time steps $t+1$ to $t+100$. We then compare the generated data against the actual 100 realized messages from the data.

\begin{figure}[htbp]
  \centering
  \begin{subfigure}{0.49\linewidth}
    \centering
    \includegraphics[width=\linewidth]{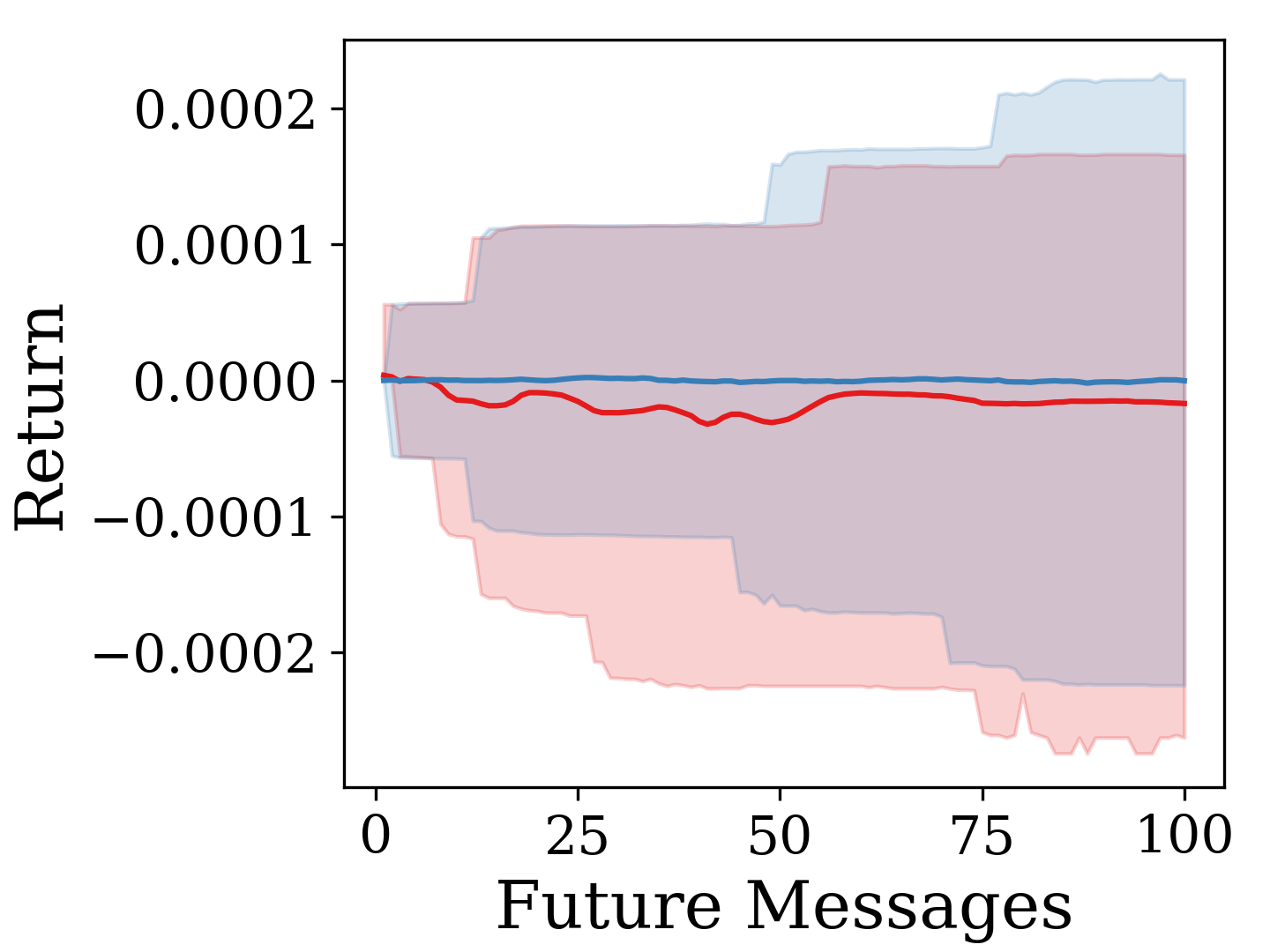}
    \caption{GOOG}
    \label{subfig:figure1}
  \end{subfigure}
  \hfill
  \begin{subfigure}{0.49\linewidth}
    \centering
    \includegraphics[width=\linewidth]{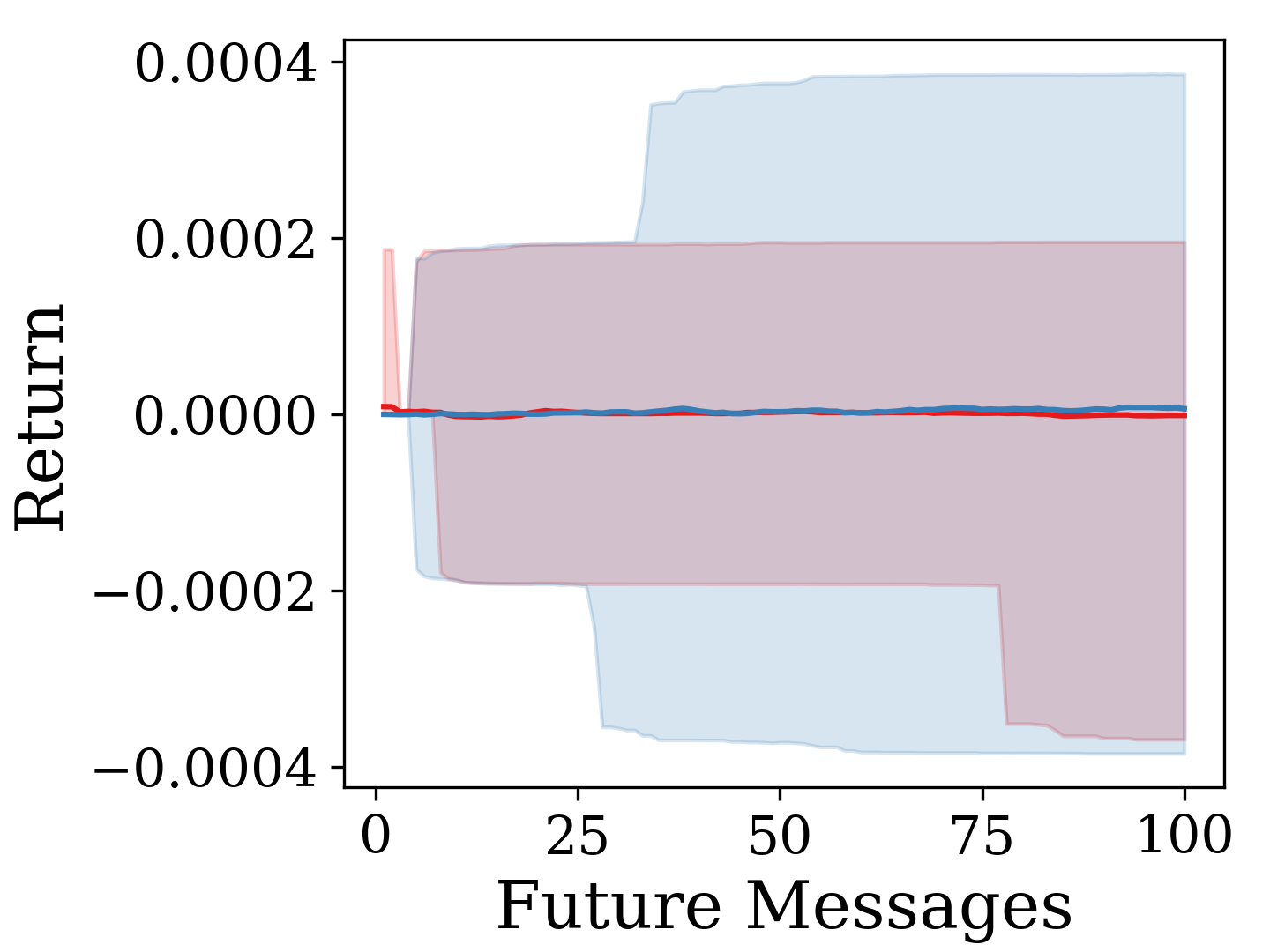}
    \caption{INTC}
    \label{subfig:figure3}
  \end{subfigure}
  \caption{
    Generated (red) and realized (blue) mid-price return distributions show how well the model matches the unconditional distribution's mean (solid lines) and 95\%ile intervals (shaded regions). Returns are calculated between the mid-price $s$ messages after, and the mid-price just before the start of generation. (n=1000 test sequences)
   }
  \label{fig:ret_comp}
\end{figure}

The mid-price $p_t$ at time $t$ is the mean of the best bid $p_s^{b(1)}$ and the best ask price $p_s^{a(1)}$, at the most recent time $s$ when there are orders on both sides of the book:
\begin{equation}
\begin{aligned}
    p_t &= \frac{p_s^{b(1)} + p_s^{a(1)}}{2}\\
    \text{where} \quad s &= \max_{u \leq t} u \quad \text{s.t.} \ p_u^{b(1)} \ \text{and} \ p_u^{a(1)} \ \text{exist}.
\end{aligned}
\end{equation}
The mid-price returns $s$ messages into the future $r_{t+s}$ between time $t$ and $t+s$ is then defined as
\begin{equation}
    r_{t+s} = \frac{r_{t+s} - r_t}{r_t}
\end{equation}
Figure \ref{fig:ret_comp} compares the generated and realized return distributions for 100 future messages. This shows that the model reproduces mid-price return distributions without this being an explicit part of the training loss. The mean returns do not exhibit any obvious drift or trend, and the shaded intervals, covering 95\% of the distribution, overlap approximately.

\begin{figure}[htbp]
  \centering
  \begin{subfigure}{0.49\linewidth}
    \centering
    \includegraphics[width=\linewidth]{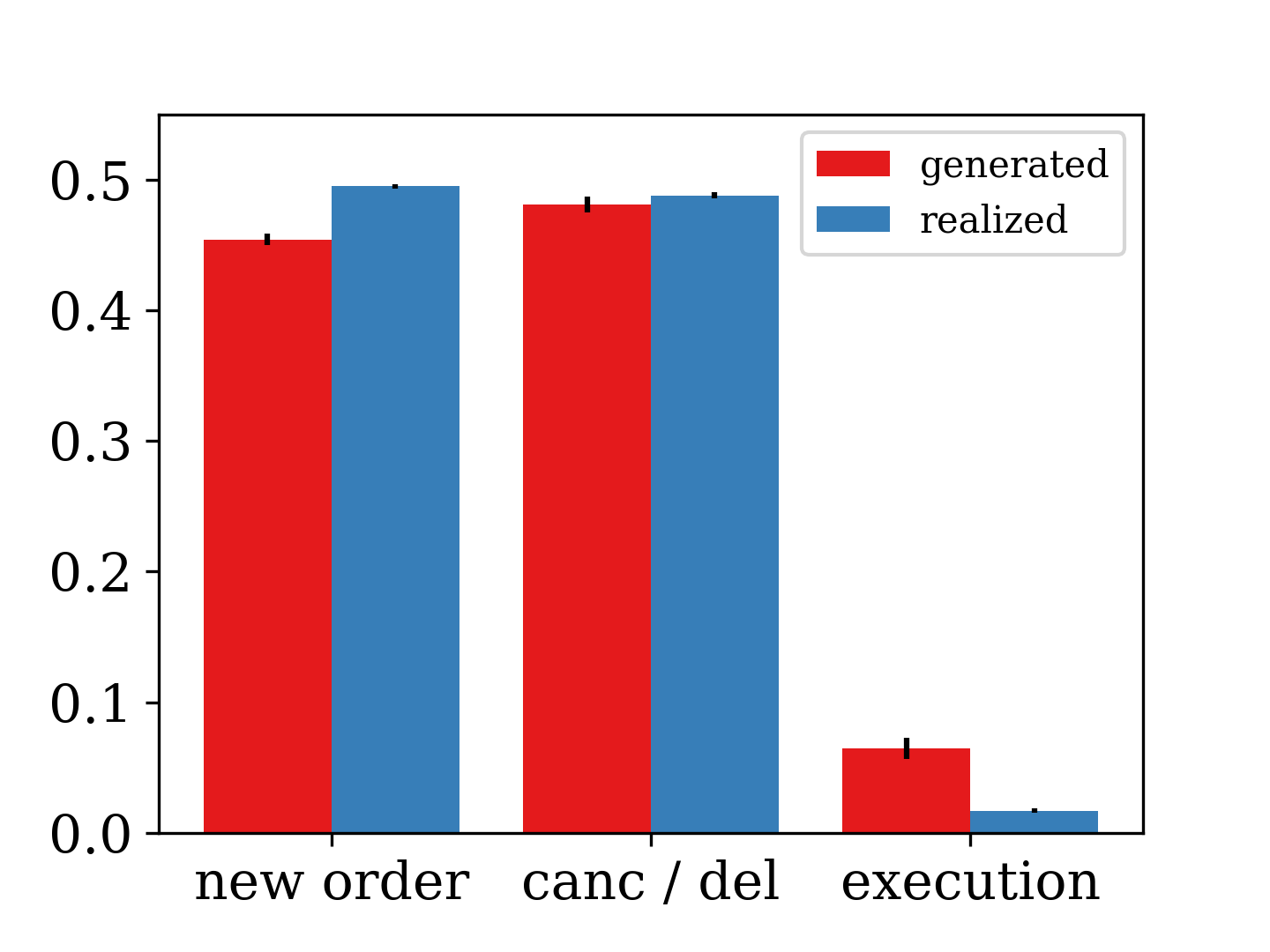}
    \caption{GOOG}
  \end{subfigure}
  \hfill
  \begin{subfigure}{0.49\linewidth}
    \centering
    \includegraphics[width=\linewidth]{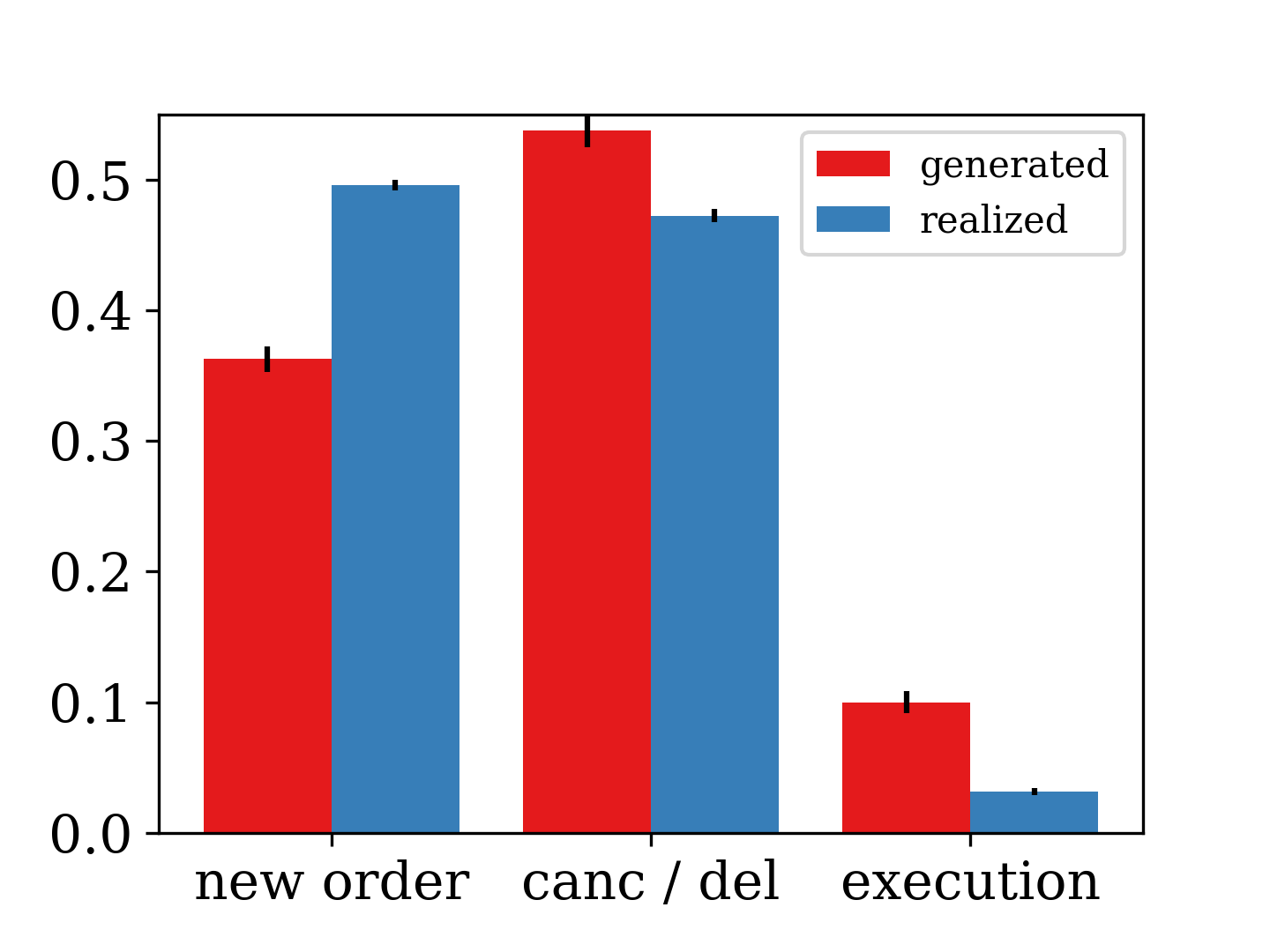}
    \caption{INTC}
  \end{subfigure}
  \caption{Comparison of order type frequency between generated (red) and realized data. While relative frequencies produced by the model roughly match magnitudes in the data, there is still a significant mismatch, with both models overestimating the probability of execution messages \emph{in test data}.
  n=(1000 sequences)$\times$(100 events) from the test set.}
  \label{fig:event_type_hist}
\end{figure}

Another desirable distribution to match is the relative frequency of message types (see Figure \ref{fig:event_type_hist}). Partial cancellations and full deletions are aggregated as the simulator only tracks individual orders if they are contained in the input sequence. As the removal of liquidity from limit orders submitted before the start of the sequence affects the aggregate initialization volume, a partial cancellation cannot be differentiated from a full deletion. While both models approximately match event frequencies, execution events are over-represented in the generated data. Similar effects can be observed for rare events in LLMs, and are potentially due to the mode-covering property of cross-entropy loss \cite{bishop2006pattern, li2023mode}.

\begin{figure}[htbp]
  \centering
  \begin{minipage}[b]{0.49\linewidth}
    \centering
    \includegraphics[width=\linewidth]{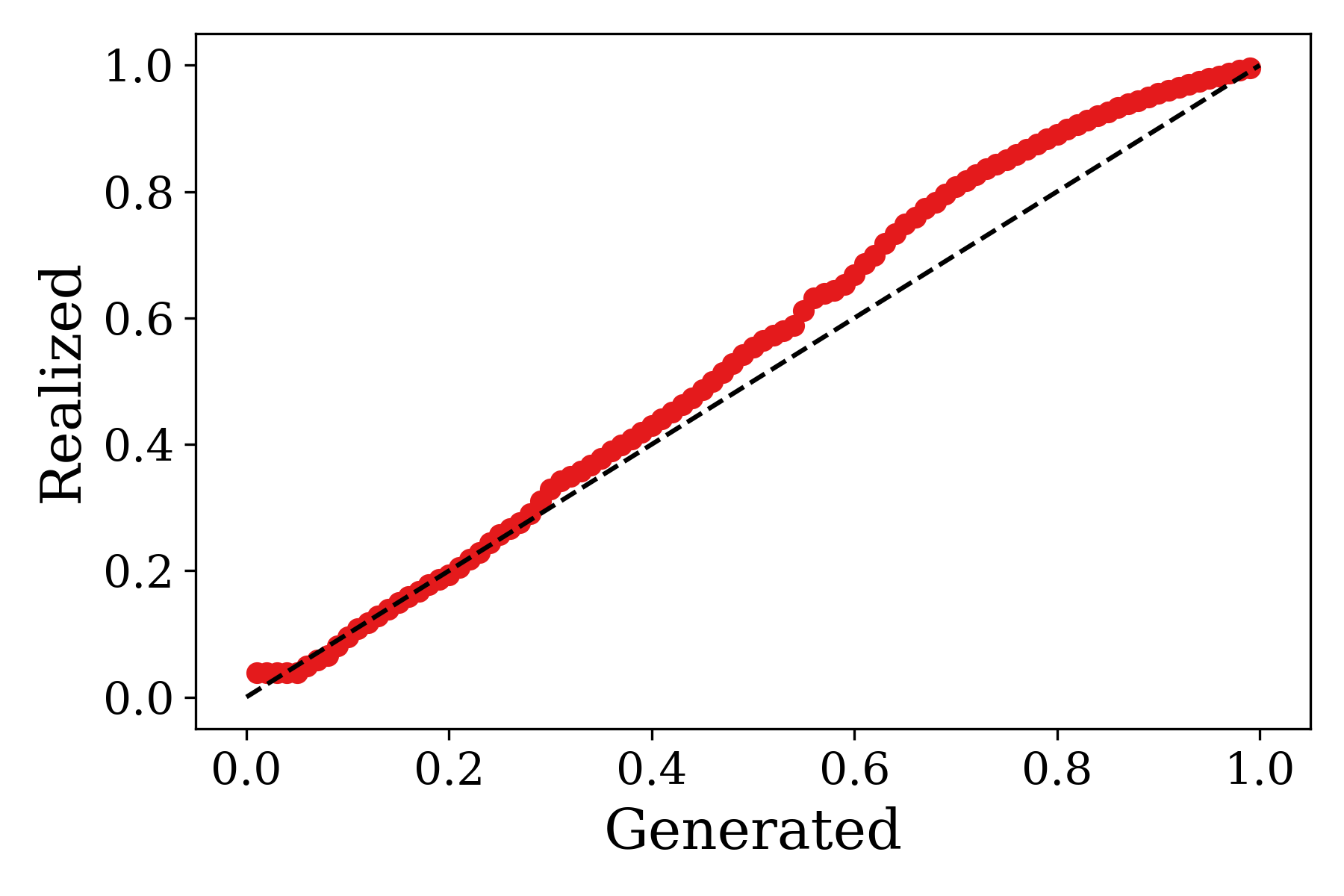}
  \end{minipage}
  \hfill
  \begin{minipage}[b]{0.49\linewidth}
    \centering
    \includegraphics[width=\linewidth]{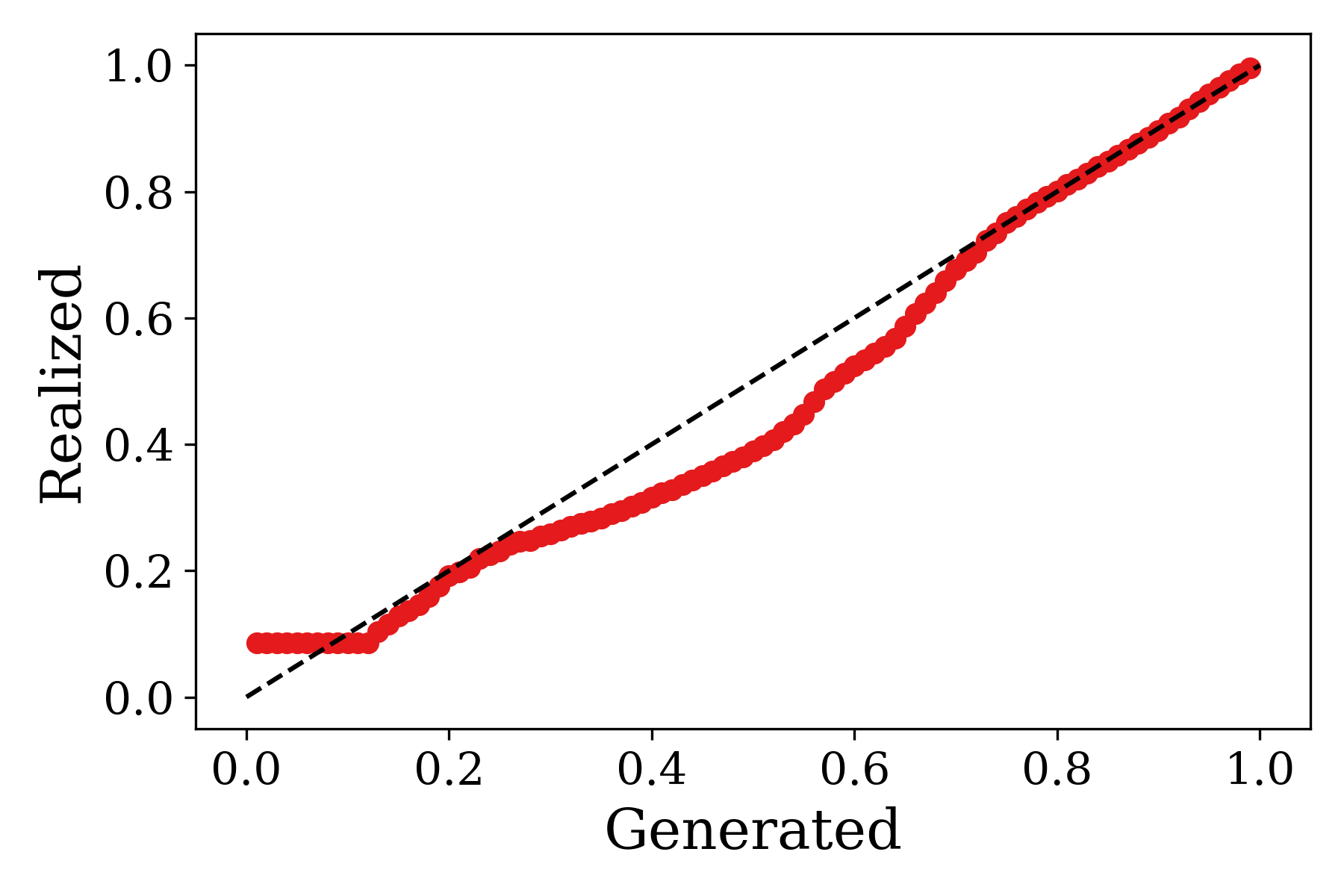}
  \end{minipage}\\
  \begin{minipage}[b]{0.49\linewidth}
    \centering
    \includegraphics[width=\linewidth]{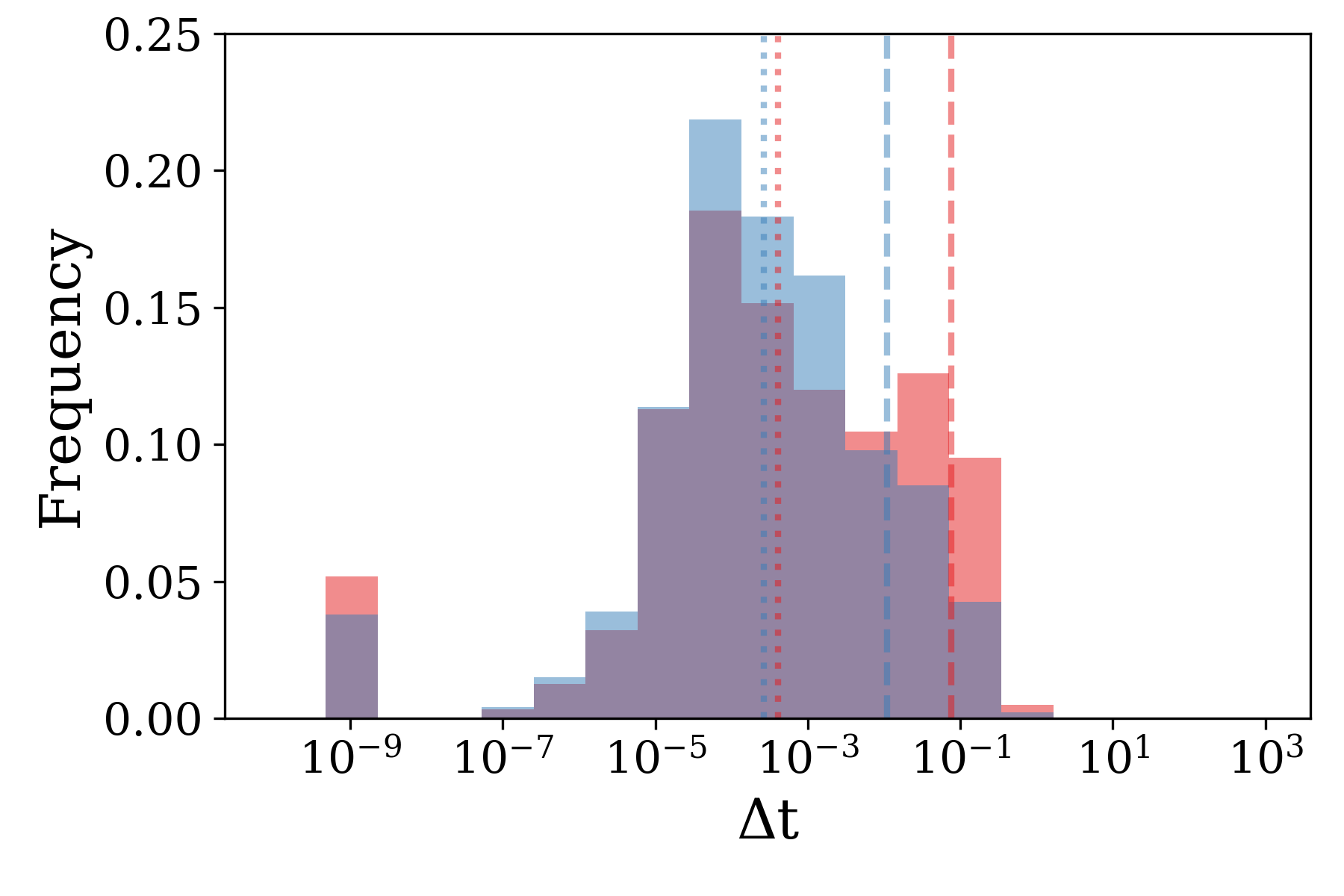}
    \subcaption{GOOG}%
  \end{minipage}
  \hfill
  \begin{minipage}[b]{0.49\linewidth}
    \centering
    \includegraphics[width=\linewidth]{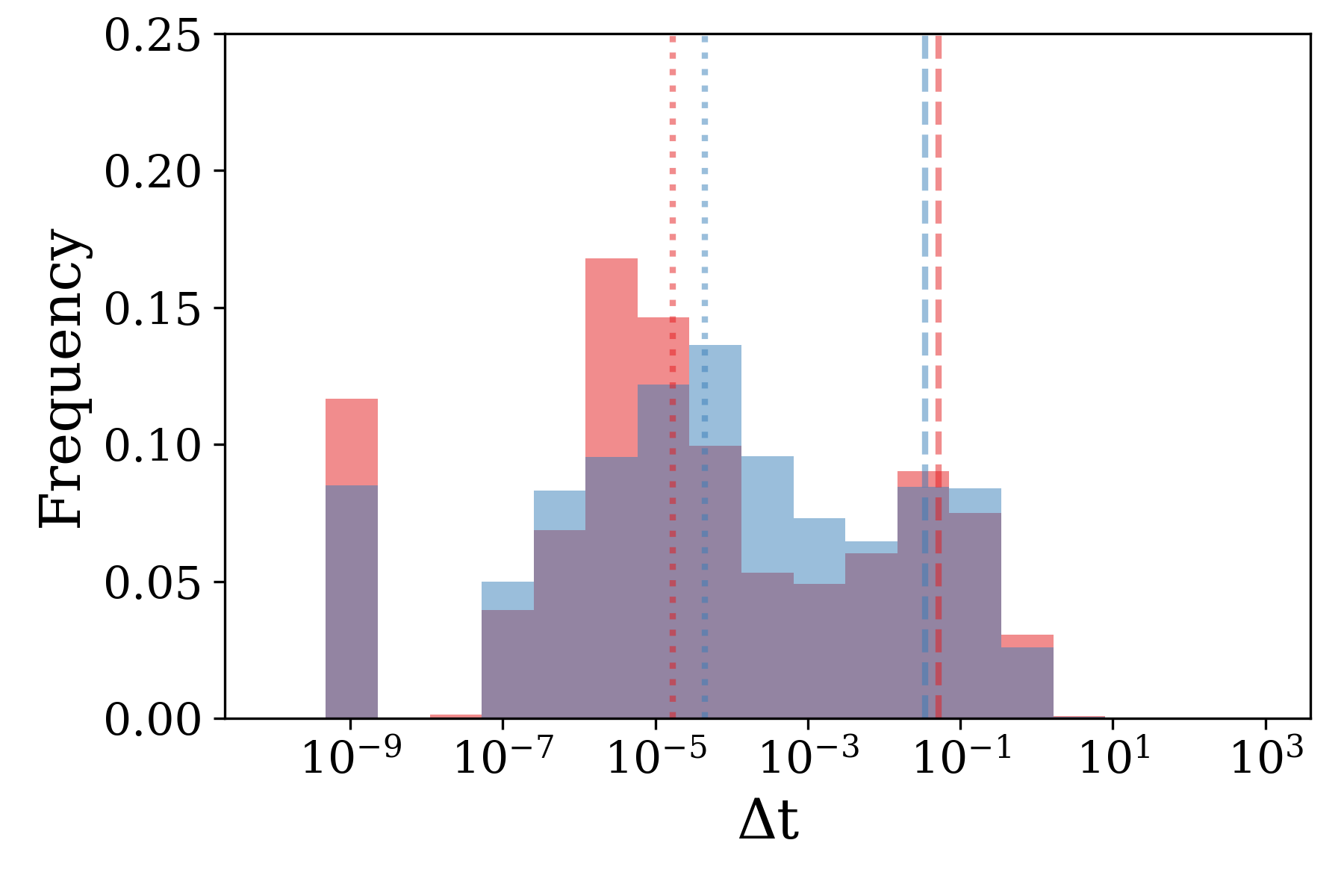}
    \subcaption{INTC}%
  \end{minipage}
  \caption{Generated decoded 4-token sequences of $\Delta t$ match message data counterparts well. First row: probability plots comparing generated and realized inter-arrival times. Second row: histograms of generated (red) and realized (blue) inter-arrival times on a log-x scale. The dashed vertical lines plot the mean and the dotted lines the median of the distributions.}
  \label{fig:delta_time_comp}
\end{figure}

As described in section \ref{sec:encoding}, for every message the model generates 4 inter-arrival time tokens $\Delta t$, which are decoded and added to the previous message's time stamp. Figure \ref{fig:delta_time_comp} compares the generated and realized message inter-arrival time distributions. The top two panels show probability plots (P-P plots), tracing the points $(F_g(\Delta t), F_r(\Delta t))$ across the support of $\Delta t$, where $F_g(\cdot)$ is the empirical cumulative distribution function (ECDF) of the generative distribution of $\Delta t$ and $F_r(\cdot)$ the realized ECDF. An equivalent interpretation is that $F_r(F_g(x))$ is shown on the y-axis for $x \in (0, 1)$, where deviations from the diagonal indicate distributional deviations. The lower panels depict corresponding histograms of $\Delta t$ on a log-x-axis, as most observations are close to 0 with a long tail. Overall the model does a good job of matching the time features.

\begin{figure}[t]
  \centering
  \begin{subfigure}{0.49\linewidth}
    \centering
    \includegraphics[width=\linewidth]{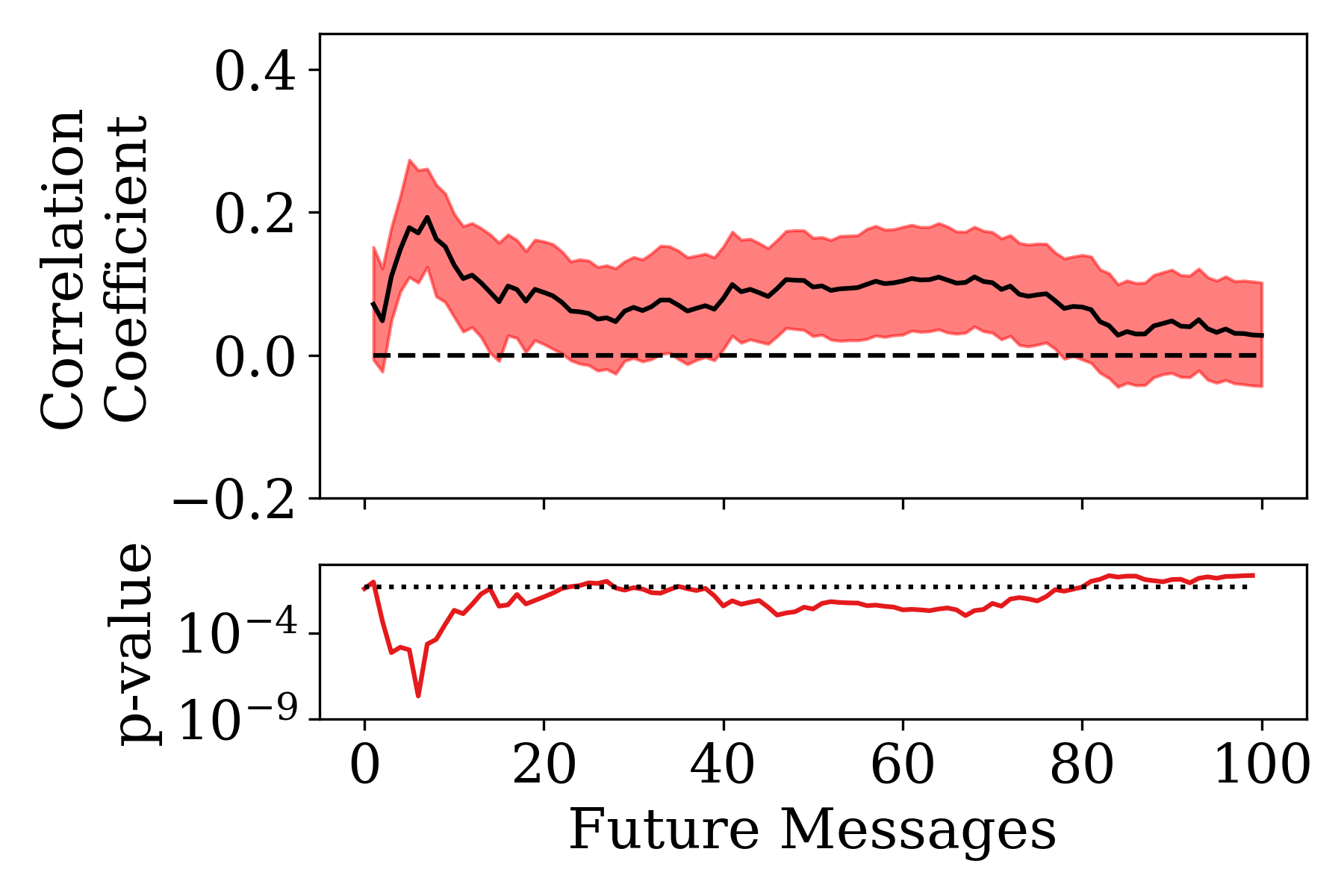}
    \caption{GOOG}
  \end{subfigure}
  \hfill
  \begin{subfigure}{0.49\linewidth}
    \centering
    \includegraphics[width=\linewidth]{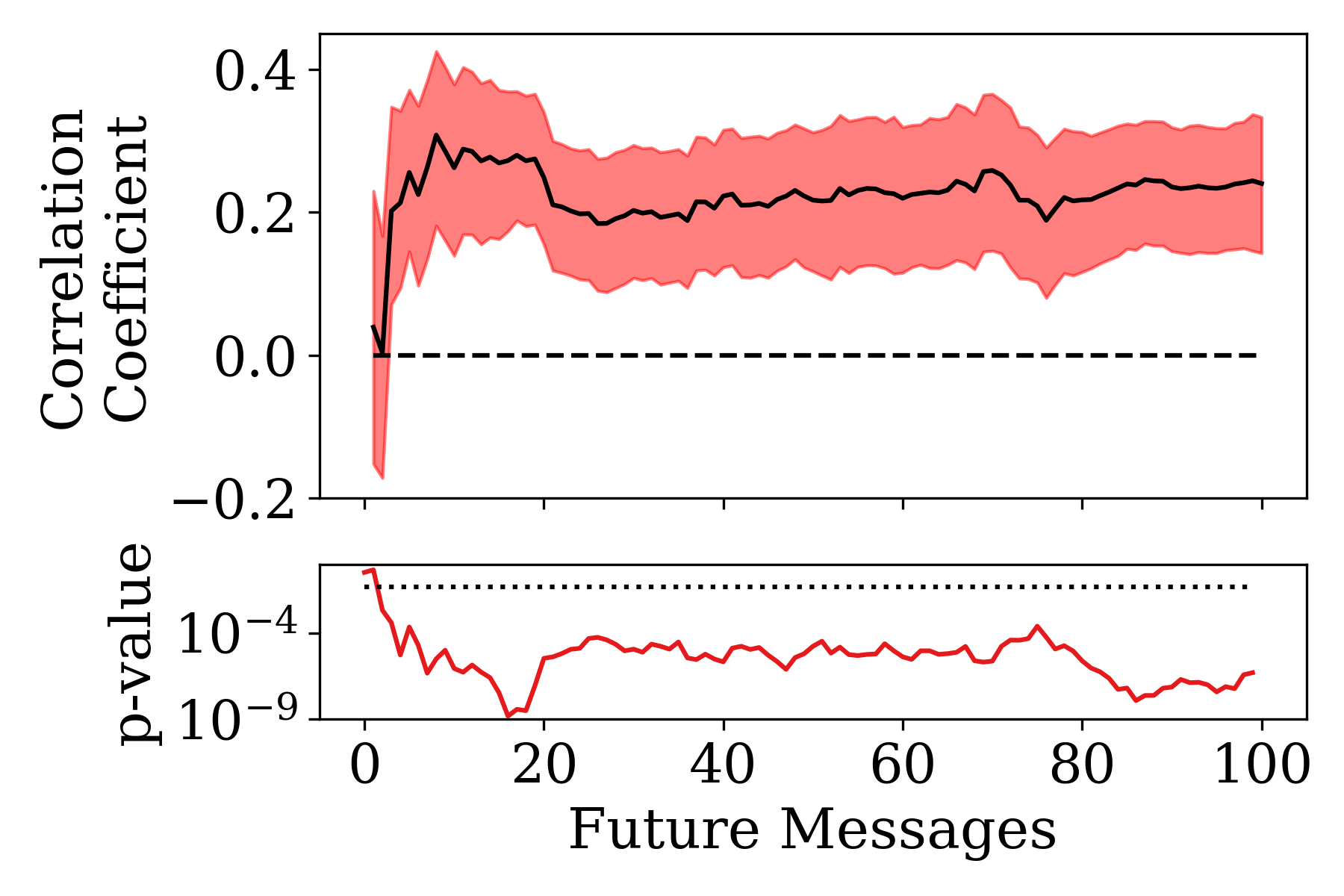}
    \caption{INTC}
  \end{subfigure}
  \caption{\emph{Top:} Pearson correlation coefficient $\rho$ between generated and realized returns, indicating directional forecasting performance. The dashed line indicates a 0 correlation. \emph{Bottom:} corresponding p-values of t-tests testing $H_1: \rho>0$ against $H_0: \rho=0$. The dotted line plots the 5\% significance threshold. The GOOG model stays below or around 5\% significance for up to 80 messages into the future, while correlation for INTC remains significant for at least 100 messages.}
  \label{fig:corr}
\end{figure}

Sampling from conditional distributions enables the model to make forecasts. To evaluate the performance of the model prediction, we calculate the Pearson correlation coefficient between the generated returns $r_{t+s}^g$ and the realized returns $r_{t+s}^r$ for $s \in [1, \ldots, 100]$ messages into the future. Figure \ref{fig:corr} shows a sustained positive correlation for both GOOG ($\rho\approx0.1$) and INTC ($\rho\approx0.2$). For INTC, the stock with less liquidity, correlations remain statistically significant at least 100 messages into the future.

\begin{figure}[htbp]
  \centering
  \begin{subfigure}{0.49\linewidth}
    \centering
    \includegraphics[width=\linewidth]{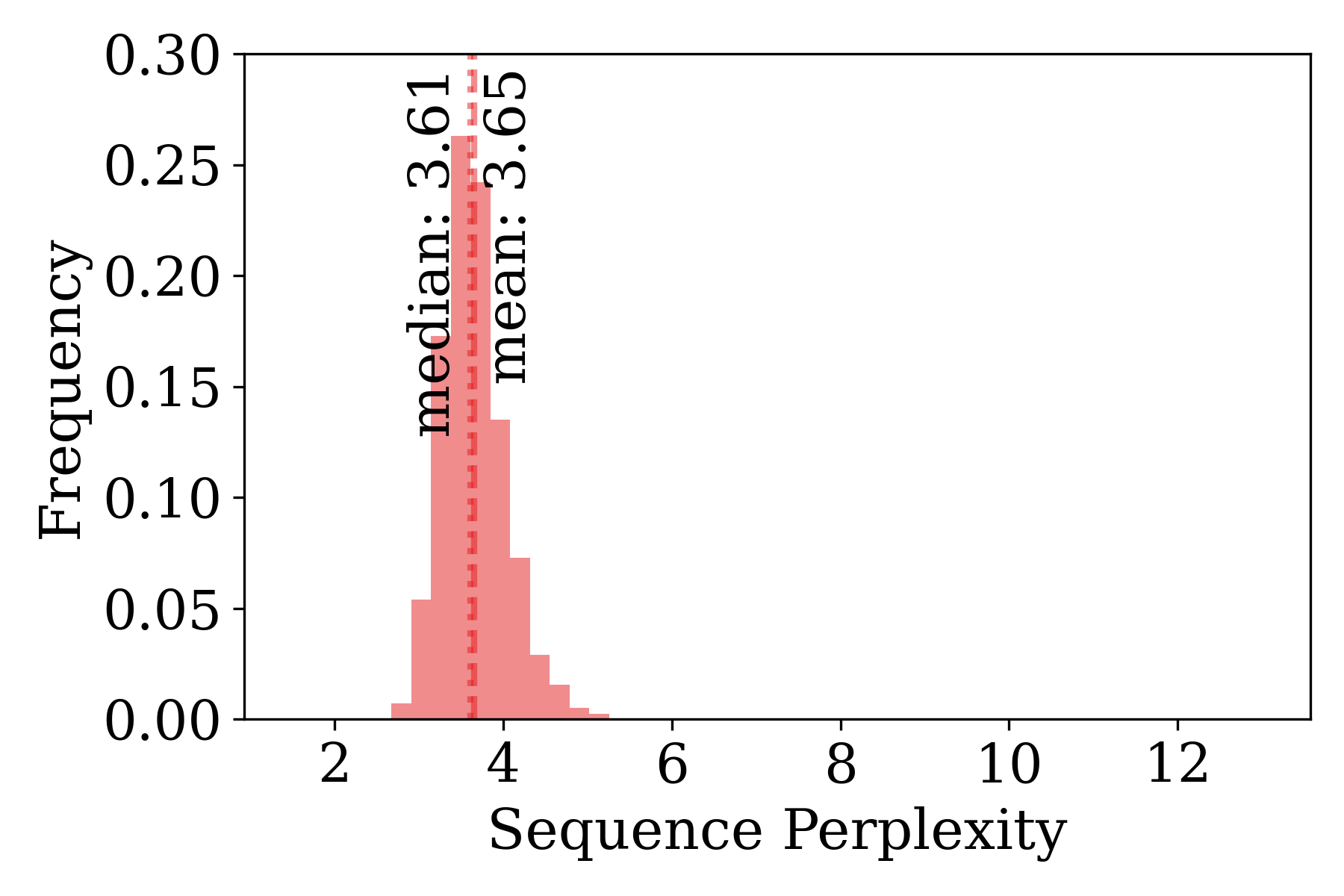}
    \caption{GOOG}
  \end{subfigure}
  \hfill
  \begin{subfigure}{0.49\linewidth}
    \centering
    \includegraphics[width=\linewidth]{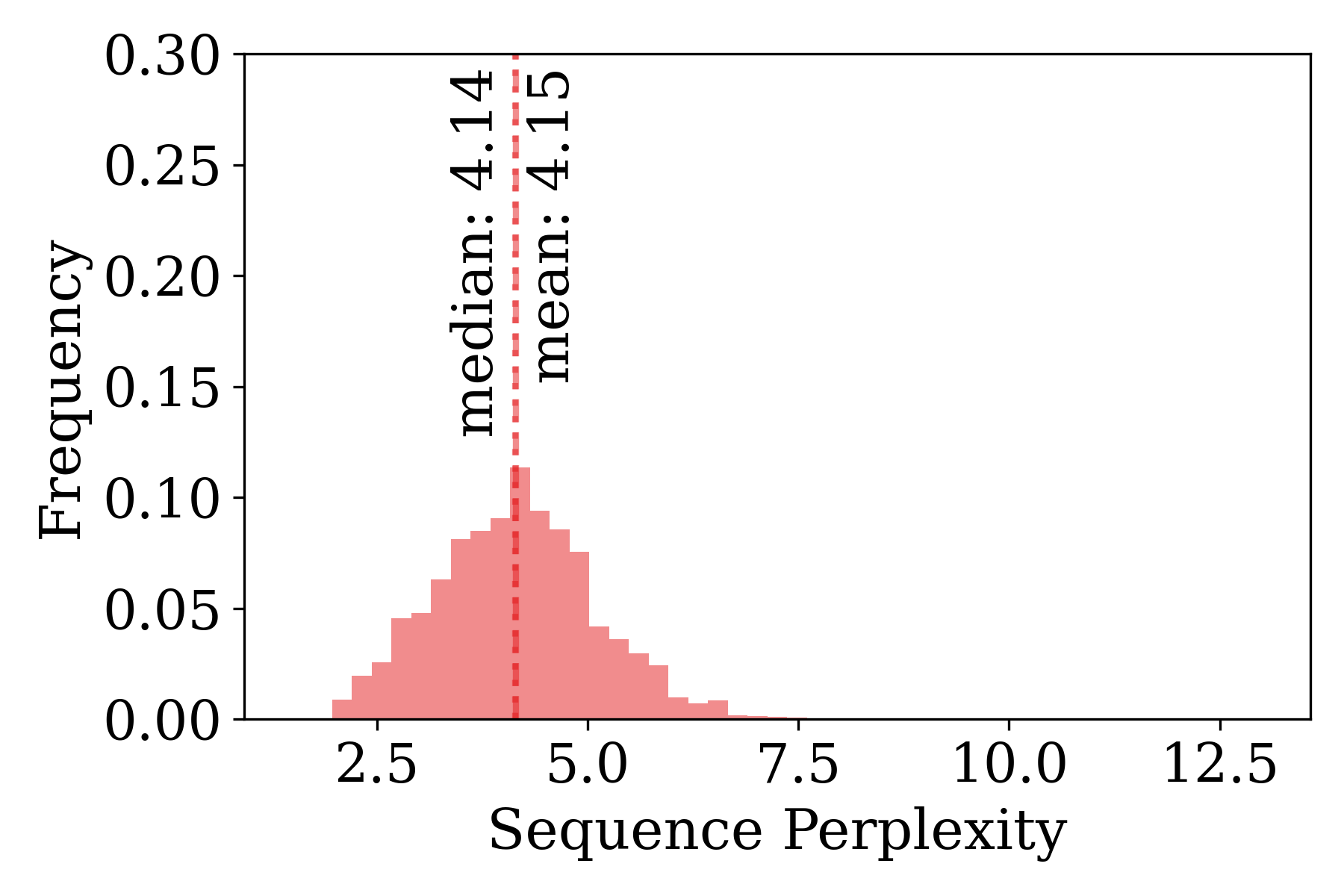}
    \caption{INTC}
  \end{subfigure}
  \caption{Per-token sequence perplexity: the GOOG model exhibits better data fit on average and fewer extreme values. Both distributions appear benign without much mass in the right tail. An observation corresponds to the PPL of a single realized 100-message sequence from the test set. Dotted lines indicate medians and dashed lines the means.}
  \label{fig:seq_perplexity_hist}
\end{figure}

Figure \ref{fig:seq_perplexity_hist} plots histograms of 100-message test sequence perplexity scores for both stocks. The model trained on GOOG data gets an overall lower perplexity than INTC with fewer extreme values of low probability sequences. This could be owed to more GOOG data being available during training as the stock has a higher trading volume and more messages in the same time period. Calculating overall per-token perplexity results in a score of $\mathbf{3.63}$ (\emph{std.err.} 0.0047) for GOOG and $\mathbf{4.04}$ (\emph{std.err.} 0.0043) for INTC. Table \ref{tab:token_perplexity} calculates the perplexity for each of the token positions generated in a message. %
We observe that later digits of inter-arrival times $\Delta t$ become harder to predict as these are higher entropy distributions. In contrast, reference time tokens are easier to predict as these usually refer to earlier time tokens in the sequence. Generally, for this reason, reference tokens have lower perplexity than their counterpart fields in the new message. Comparing GOOG with INTC we observe, that the GOOG model has higher perplexity in generating an order's price level, while INTC has higher uncertainty of order sizes. The difference in price level predictability agrees with the fact that INTC, as a small-tick stock, exhibits a denser order book on average \cite{eisler2012price}, while GOOG order sizes have lower entropy than INTC.

\begin{table*}[tbp]
  \small
  \centering
  \caption{Token Perplexity Scores by Message Position}
  \label{tab:token_perplexity}
  \adjustbox{max width=\textwidth}{
    \begin{tabular}{@{}l*{17}{c}@{}}
      \toprule
      \textbf{Stock} & \multicolumn{1}{c|}{Type} & \multicolumn{1}{c|}{Direction} & \multicolumn{2}{c|}{Price} & \multicolumn{1}{c|}{Size} & \multicolumn{4}{c|}{$\Delta$ t} & \multicolumn{2}{c|}{Price\textsubscript{ref}} & \multicolumn{1}{c|}{Size\textsubscript{ref}} & \multicolumn{5}{c}{Time\textsubscript{ref}} \\
      \midrule
      GOOG & 2.15 & 1.71 & 1.04 & 5.55 & 3.18 & 1.00 & 6.21 & 405.49 & 770.01 & 1.04 & 1.45 & 1.07 & 1.01 & 1.63 & 2.41 & 2.09 & 1.93 \\
      INTC & 1.92 & 1.55 &  1.02  & 2.41 & 13.51  & 1.01 &  5.45 & 122.93 & 524.66 & 1.03 & 1.21 & 1.22 & 1.02 & 1.96 & 3.76 & 6.58 & 7.94 \\
      \bottomrule
    \end{tabular}
  }
\end{table*}

\section{Conclusions} \label{sec:conclusion}

We develop an end-to-end autoregressive generative model for electronic exchange message flow, trained on tokenized message sequences and LOB states, using a custom tokenizer. By embedding the model in an inference loop with a market replay simulator, we are able to generate full-fidelity granular trajectories of entire LOBs.
A challenge of predictive machine learning models in finance, particularly in high-frequency market micro-structure, is market impact, i.e. the fact that market actions affect dynamics. Modeling realistic market impact is thus an important open problem, for which generative models of this kind provide a promising solution.

For tasks in natural language processing, autoregressive LLMs have proven superior to GAN-based models in cases where sufficient training data is available. Arguably, market micro-structure provides a similar domain with 2500 different stocks trading on the NASDAQ exchange alone, each with up to millions of daily messages. Our results show that such models are reasonable, can provide good performance, and could potentially be scaled up to something like autoregressive large financial environment models, which could be used for reinforcement learning. Similar approaches using GANs have not yet been very successful due to agents learning to exploit model errors \cite{coletta2023conditional}. 

One challenge to generative micro-structure models is posed by message-arrival times. To our knowledge, we propose the first generative micro-structure data model, which treats times the same as other data features and learns a generative distribution, compared to e.g. binning time intervals as done in \cite{hultin2023generative}.

Despite the fact that the model is trained on tokenized micro-structure messages, we are able to match data distributions, such as the distribution of mid-price returns. 
This radical bottom-up approach, in combination with large neural networks, has been driving recent successes in LLMs and could similarly usher in the next generation of generative financial models. 

Given these promising results, our work provides many interesting directions for future research. Current model limitations include errors when generating referential orders and computationally intensive message error correction. These issues could be abated by increasing the model and data set size. Besides larger models, future work could similarly investigate training on longer sequences. Different architectural choices, either improvements to the S5 layer \cite{zhang2022effectively}, or efficient transformer-based networks \cite{child2019generating} could be explored as alternatives.

\begin{acks}
We thank Jan-Peter Calliess and Chris Lu for interesting discussions at different stages of this project.
\end{acks}

\bibliographystyle{ACM-Reference-Format}
\bibliography{sample-base}

\end{document}